\documentclass{aa}
\usepackage[varg]{txfonts}

\usepackage{array, multirow, graphicx,natbib,txfonts,color,tabularx,amsmath}
\usepackage{lscape,float,placeins}
\usepackage{threeparttable}
\bibpunct{(}{)}{;}{a}{}{,}

\newcommand{\FeH}{\left[\mbox{Fe/H}\right]}
\newcommand{\XY}{\left[\mbox{X/Y}\right]}
\newcommand{\XFe}{$\left[\mbox{X/Fe}\right]$}

\newcommand{\MgFe}{$\left[\mbox{Mg/Fe}\right]$}
\newcommand{\AlFe}{$\left[\mbox{Al/Fe}\right]$}
\newcommand{\NaFe}{$\left[\mbox{Na/Fe}\right]$}
\newcommand{\SiFe}{$\left[\mbox{Si/Fe}\right]$}
\newcommand{\kms}{km\,s$^{-1}$}
\newcommand{\Teff}{$T_{\mbox{{\scriptsize eff}}}$}
\newcommand{\logg}{$\log g$}

\newcommand{\vmic}{$v_{\mbox{{\scriptsize mic}}}$}
\newcommand{\abund}[1]{\log\,\varepsilon(\text{#1})}
\newcommand{\ch}[1]{\multicolumn 1 c {#1}}
\begin{document}

\title{Atomic Diffusion and Mixing in Old Stars V: A deeper look into the Globular Cluster NGC\,6752 \thanks{Based on data collected at the ESO telescopes under programs 079.D-0645(A) and 081.D-0253(A)}\fnmsep\thanks{Tables 2 and 8 are only available in electronic form at the CDS via anonymous ftp to cdsarc.u-strasbg.fr (130.79.128.5) or via http://cdsweb.u-strasbg.fr/cgi-bin/qcat?J/A+A/???/???}}

\author{Pieter Gruyters \and Thomas Nordlander  \and Andreas J. Korn }

\offprints{pieter.gruyters@physics.uu.se}

\institute{Division of Astronomy and Space Physics, Department of Physics and Astronomy, Uppsala University, Box 516, 75120 Uppsala, Sweden  }

\date{Received 6 February 2014 / Accepted 21 May 2014}

\authorrunning{P. Gruyters}
\titlerunning{Atomic Diffusion in NGC6752}

%*****************************************************************************
%                      ABSTRACT
%*****************************************************************************
\abstract
%context
{Abundance trends in heavier elements with evolutionary phase have been shown to exist in the globular cluster NGC\,6752 $(\FeH = -1.6)$. These trends are a result of atomic diffusion and additional (non-convective) mixing. Studying such trends can provide us with important constraints on the extent to which diffusion modifies the internal structure and surface abundances of solar-type, metal-poor stars.}
%aims
{Taking advantage of a larger data sample, we investigate the reality and the size of these abundance trends and address questions and potential biases associated with the various stellar populations that make up NGC\,6752.
}
%Methods 
{We perform an abundance analysis by combining photometric and spectroscopic data of 194 stars located between the turnoff point and the base of the red giant branch. Stellar parameters are derived from $uvby$ Str\"omgren photometry. Using the quantitative-spectroscopy package SME, stellar surface abundances for light elements such as Li, Na, Mg, Al, and Si as well as heavier elements such as Ca, Ti, and Fe are derived in an automated way by fitting synthetic spectra to individual lines in the stellar spectra, obtained with the VLT/FLAMES-GIRAFFE spectrograph.}
%Results
{Based on $uvby$ Str\"omgren photometry, we are able to separate three stellar populations in NGC\,6752 along the evolutionary sequence from the base of the red giant branch down to the turnoff point. We find weak systematic abundance trends with evolutionary phase for Ca, Ti, and Fe which are best explained by stellar-structure models including atomic diffusion with efficient additional mixing. We derive a new value for the initial lithium abundance of NGC\,6752 after correcting for the effect of atomic diffusion and additional mixing which falls slightly below the predicted standard BBN value.}
%Conclusion
{We find three stellar populations by combining photometric and spectroscopic data of 194 stars in the globular cluster NGC\,6752. Abundance trends for groups of elements, differently affected by atomic diffusion and additional mixing, are identified. Although the statistical significance of the individual trends is weak, they all support the notion that atomic diffusion is operational along the evolutionary sequence of NGC\,6752. }

% Max 6 keywords!
\keywords{stars: abundances - stars: atmospheres - stars: fundamental parameters - globular cluster and associations: NGC\,6752 - techniques: spectroscopic }

\maketitle

%******************************************************************
%                    INTRODUCTION
%******************************************************************

\section{Introduction}\label{sect:intro}
Stellar spectroscopy of stars with spectral types F, G and K is a mature field of astrophysics. At least for simple atoms like lithium and sodium, the matter-light interaction taking place in stellar atmospheres can nowadays be modelled from first principles, with due consideration of hydrodynamics and departures from local thermodynamic equilibrium (LTE). Using long-lived stars as data carriers of different phases of Galactic evolution, much has been learned about the matter cycle and the general build-up of the Galactic inventory of chemical elements with time. However, our inferences are not free from uncertainties and limitations. One such inquietude is the theoretical expect that the surface abundances of late-type stars with shallow outer convection zones are a direct function of time, due to element-separating effects collectively referred to as atomic diffusion \citep{Michaud1984}. Subject to the prevailing forces (gravity, radiation pressure), chemical elements will be pushed into/out of the atmosphere, leaving the visible surface somewhat enriched/depleted, to various degrees for different elements. Given that convection can efficiently counteract atomic diffusion, the effects are expected to lead to sizable surface effects only in hotter stars, i.e. in the context of this paper the F-type turnoff-point stars. During prior and subsequent evolution, the deep outer convection zone retains/restores the surface abundances (with the notable exception of lithium which is mostly destroyed during the course of the first dredge-up on the red-giant branch).

In this series of papers, we investigate to which extent careful spectroscopic studies can trace small yet systematic abundance differences between groups of stars representing different phases in the evolution of globular-cluster stars. The basic idea is simple: with sufficiently accurate spectroscopic modelling, inferred abundance differences can be compared to the predictions of stellar-structure models including the effects of atomic diffusion.

Earlier papers in the series have traced such effects in NGC\,6397 at $\FeH \footnote{We adopt here the usual spectroscopic notations that $\XY \equiv \log\,(N_{\rm X}/N_{\rm Y})_{*} - \log\,(N_{\rm X}/N_{\rm Y})_{\odot}$, and that $\log\,\varepsilon({\rm X})\equiv \log\,(N_{\rm X}/N_{\rm H})+12$ for elements X and Y. We assume also that metallicity is equivalent to the stellar [Fe/H] value.} = -2.1$ \citep{Korn2007, Lind2008, Nordlander2012} and NGC\,6752 at $\FeH = -1.6$ \citep[][hereafter Paper~I]{Gruyters2013}. It was shown that the turnoff-point (TOP) stars do show systematically lower abundances (by 0.05-0.2\,dex) than the red-giant branch (RGB) stars, with intermediate groups falling in between.  But the predictions of atomic diffusion are not supported by these observations straight off. To achieve agreement between theory and observation, atomic diffusion needs to be counteracted by what is referred to as additional mixing, i.e.\ non-canonical mixing not provided by convection in the mixing-length formulation. Some mixing beyond the formal extent of the convection zone in the turnoff-point phase of evolution is also needed to explain the properties of the Spite plateau of lithium \citep{Spite1982} which at a given metallicity is found to be mono-valued and not a function of effective temperature \citep{Richard2005}. Such mixing can be prescribed in stellar-structure models by an extra term in the diffusion equation, adding a simple analytical function that accomplishes mixing down to layers of a specific temperature. The free parameter involved in this modelling, the efficiency relative to atomic diffusion, has to, at present, be determined empirically from the observations, i.e.\ from the amplitude of the elemental abundance trends.

For stars in NGC\,6397, the efficiency of additional mixing is well-constrained and falls within the range that prevents lithium destruction which would be incompatible with the properties of the Spite plateau (see above). The agreement between theory and observation is very good and succeeds in describing abundance structures seen for the first time in lithium and calcium \citep{Korn2006, Lind2009}. There is, however, no complete consensus on the effective-temperature scale to be applied, with consequences for the abundance trends of e.g.\ lithium between main-sequence (MS) and sub-giant-branch (SGB) stars \citep{Gonzalez2009b}.

For stars in NGC\,6752, the results are less compelling, mainly because the abundance trends are weaker in this cluster. Nonetheless, with somewhat more efficient additional mixing, the trends can be reproduced and the diffusion-corrected lithium abundance raised to levels compatible with WMAP-calibrated predictions of standard Big-Bang Nucleosynthesis \citep{Gruyters2013}.\\

In this paper, we revisit NGC\,6752 with additional observations obtained with the GIRAFFE spectrograph at the VLT. With improved stellar statistics, we can address questions and potential biases associated with the various stellar populations that make up NGC\,6752. The paper is organized in the following manner: Section 2 describes the observations, both photometric and spectroscopic, and the data reduction providing the material for the analysis presented in  Section 3. Section 4 describes the results, while Section 5 discusses their scientific significance. In Section 6, the results are summarized and some conclusions are drawn.

%#######################################################################################

%******************************************************************
%                    OBSERVATIONS
%******************************************************************

\section{Observations and Data Reduction}\label{sect:obs}

\subsection{Photometry}
Photometric data was obtained using the DFOSC instrument on the 1.5\,m telescope on La Silla, Chile, in 1997 and consist of $uvby$ Str\"omgren photometry. The data is the same as used in Paper~I, and details on the data reduction and photometric calibration procedures can be found in \citet{Grundahl1998,Grundahl1999}. 

\begin{figure}
\begin{center}
\includegraphics{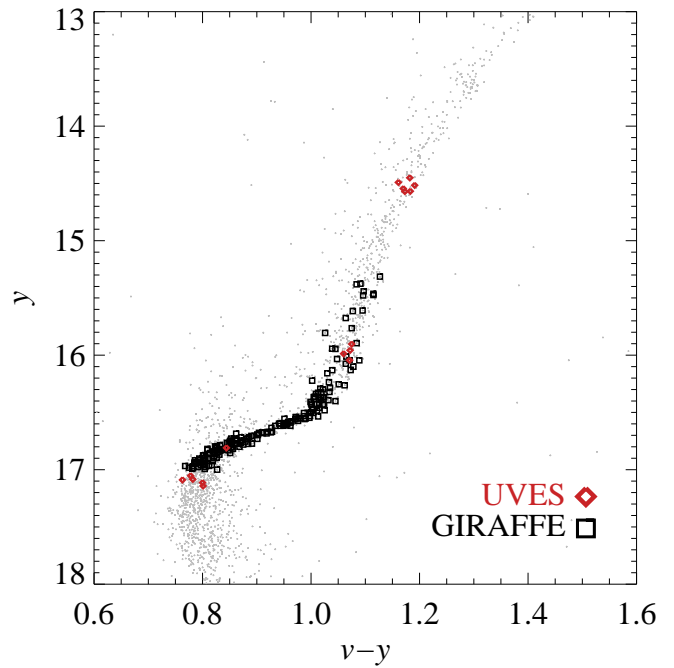}
\caption{Observed colour-magnitude diagram of NGC 6752. The spectroscopic targets are marked by black squares for the GIRAFFE sample. For comparison we have included the UVES targets from \citet{Gruyters2013} as red diamonds.}\label{Fig:CMD}
\end{center}
\end{figure}

The sample used for spectroscopic analysis consists of 194 stars located between the cluster turnoff point (TOP) at $V=17.0$ and the base of the red giant branch (bRGB) up to $V=15.3$. A colour-magnitude diagram (CMD) of the total sample is given in Fig.~\ref{Fig:CMD}. We construct colour-magnitude fiducial sequences by averaging the colours of cluster members along the sequence. By interpolating colours at constant $V$-magnitude onto this sequence, we effectively remove the observational scatter, with residual errors dominated by uncertainties in the shape of the fiducial sequence. The best-fitting fiducial is found by trial and error. We find the most reliable method of constructing colour-magnitude fiducials to be quadratic interpolation between points placed manually such that the resulting residuals along the sequence were minimized. Finally, we apply a reddening correction of $E(B-V) = 0.04$  \citep[latest web update]{Harris1996} with the relation coefficients given by \citet{Ramirez2005}, and $E(v-y)=1.7\times E(b-y)$.

\subsection{Spectroscopy}
The spectroscopic observations were obtained in Service Mode during ESO period 79 (Program ID: 079.D-0645(A)) and period 81 (Program ID: 081.D-0253(A)) using the multi-object spectrograph FLAMES/GIRAFFE mounted on the ESO VLT-UT2 Kueyen \citep{Pasquini2003}. Running in MEDUSA mode, FLAMES-GIRAFFE can simultaneously observe 132 objects. On average we dedicated 15 of the 132 fibres to a simultaneous monitoring of the sky background. As our faintest targets reach $V \approx 17$, we typically coadded multiple long exposures (up to $4215\,\text s$) in order to reach spectra of sufficient quality with signal-to-noise ratios per pixel $(S/N)$ above 30. Including overhead, we spent a total of $70$\,h under good seeing conditions ($\sim 0.7''$) during grey-to-bright lunar time periods using six high-resolution ``B''-settings (HR9B, HR10B, HR11B, HR13B, HR14B, and HR15B). 

\begin{table*}
\caption{Observational data of the NGC\,6752 stars observed with FLAMES-GIRAFFE.}
\label{Tab:log}      
\centering          
\begin{tabular}{lcccccc} 
\hline\hline
 Setting & Wavelength & Resolution & $t_{\mbox{\scriptsize exp}}$ & $N_\text{stars}$ & ESO period & S/N  \\
 \# & range (\AA) & $\lambda/\Delta\lambda$ & (hr) & & & Min-Median-Max \\ 
\hline                    
 HR9 & 5143-5356 & 25900 & 12.9 & 98 & P81 & 38-65-134 \\
 HR10 & 5339-5619 & 19800 & 3.5 & 99 & P81 & 26-53-84 \\
 HR11 & 5597-5840 & 24200 & 8.4 & 138 & P79 \& P81 & 18-61-104 \\ 
 HR15 & 6607-6965 & 19300 & 4.5 & 126 & P79 & 30-87-142 \\  
 \hline
\end{tabular}
\tablefoot{Only data analysed in this work are listed.}
\end{table*}

In this work we present the data of four of the six settings, covering a combined spectral range of 1100\,\AA. Stars in different evolutionary stages are distributed randomly across the field-of-view. We find no gradient in background light over the field and subtract a single, median averaged sky-spectrum from the stellar spectra obtained during each observation. The setup is summarised in Table~\ref{Tab:log} for the subset of spectra analysed in this work.

As UVES targets were prioritized during observations, fibre positioning limitations resulted in only 37 stars out of the full sample of 194 being observed in all six settings. The largest overlapping sample consists of 105 stars observed in HR11B covering e.g. the \ion{Mg} i 5711\,\AA\ line and the \ion{Na} i 5682-5688\,\AA\ doublet, and HR15B covering the \ion{Li} i 6707\,\AA\ line and the \ion{Al} i 6696-6698\,\AA\ doublet. Additionally, 99 stars were observed in both HR10B and HR11B, the former covering the \ion{Mg} i 5528\,\AA\ line.

We executed basic data reduction on the spectroscopic data using the GIRAFFE pipeline maintained by ESO. 
Individual exposures, ranging between 1 and 11 per star, were coadded after sky-subtraction, normalization, radial-velocity correction and cosmic-ray rejection. As we have only a single exposure for some stars, we opted for a threshold-rejection method of affected pixels and their neighbours, and visually inspected all spectra before and after coaddition.

The minimum, maximum and median $S/N$-ratios per rebinned pixel of the coadded spectra are given in Table~\ref{Tab:log} for the different settings. These values are computed for each spectral order in the reasonably line-free continuum regions used for normalization, and are subsequently applied as flux error estimates.

%#######################################################################################

%******************************************************************
%                                  ANALYSIS
%******************************************************************

\section{Analysis}\label{sect:analysis}
\subsection{Effective temperature}\label{subsect:phot}
A photometric effective temperature (\Teff) scale was constructed from the $(v-y)$ and $(b-y)$ colour indices using relations for main sequence stars \citep{Alonso1996,Korn2007} and giant stars \citep{Alonso1999,AlonsoEratum, Korn2007}, assuming $\FeH =-1.60$ for all stars. Both relations are calibrated on the infrared flux method. As our sample stretches from the TOP at the very end of the MS to the bRGB, we evaluate separately the dwarf and giant calibrations, and interpolate \Teff\ linearly as a function of \logg\ between them.

Comparing the $(v-y)$ and $(b-y)$ calibrations, differences are rather small: the former is hotter by $16 \pm 57$\,K for the dwarf and $6 \pm 55$\,K for the giant calibration. The difference is at most roughly 80\,K in a narrow part of the sub-giant branch, which falls outside the colour ranges covered by the calibrations. Temperatures based on $(v-y)$ and $(b-y)$ exhibit a scatter about the fiducial sequences of 40\,K and 80\,K, respectively. Comparing the fiducial sequences on the two colour indices, we find a dispersion of just 23\,K, including the aforementioned biases.

While the $v$-band is in principle sensitive to variations in chemical composition, i.e. different CN band strengths \citep{Carretta2011}, our fiducial sequence does not appear to be strongly affected by this. We judge potential errors in placement of the fiducial sequence due to such effects to be at most $+20$\,K, for our coolest stars where the sensitivity is greatest. We opt to base our temperature scale on the $(v-y)$ relation as it exhibits less scatter, higher sensitivity on the RGB, and creating its fiducial appears to be less susceptible to systematic errors. Taking into account the statistical uncertainties present, we adopt a representative uncertainty of 40\,K in \Teff.

The calibration of \citet{Casagrande2010} on the $(b-y)$ index produces higher temperatures by 120--150\,K, with a slight dependence on colour. No corresponding calibration is available for $(v-y)$. With the main difference to previous calibrations being in the zero-point (as indicated in that paper), we may assume a similar absolute uncertainty would apply to the $(v-y)$ index. \\

\subsection{Surface gravity}\label{subsect:grav}
We employ photometric surface gravities, assuming a distance modulus of 13.13 \citep[latest web update]{Harris1996}. The bolometric corrections of \citet{Alonso1999,AlonsoEratum} are applied to observed $V$-magnitudes, as a function of metallicity and \Teff, to compute luminosities. Stellar masses are interpolated directly as a function of \Teff\ from a 13.5\,Gyr isochrone \citep[from][]{Gruyters2013}, and range from 0.79 to 0.815\,M$_\odot$. The resulting values of \logg\ range from 4.0 at \Teff\ $=6050$\,K (TOP) to 3.0 at \Teff\ $=5160$\,K (bRGB). The final stellar parameters, along with observational parameters, are given in Table\,\ref{Tab:SPs}.

The surface gravities are most sensitive to uncertainties in \Teff, where a change of $+100$\,K propagates into \logg\ changes of $+0.04$\,dex on the TOP, or $+0.03$\,dex on the bRGB. An effect of $\delta \log g=+0.01$\,dex would require, e.g., an increase in stellar mass of 0.02\,M$_\odot$, or an increase in $V$-magnitude by 0.025\,mag, be it due to observational uncertainties or errors in distance modulus or bolometric correction. \citet{Sbordone2011} examined the effects of chemical composition variations on isochrones and photometry. Concerning isochrones, they found that only extreme changes in nitrogen or helium abundance could produce any discernible effects. \citet{Cassisi2013} expanded on this work and note a composition-driven increase in \Teff\ by $+30$\,K and $\Delta \log(L/L_\odot) \approx 0.017$, where partial cancellation results in an effect $\delta \log g = +0.01$\,dex. On the bRGB on the other hand, $\delta$\Teff\ $\approx 0$\,K, giving an 0.01\,dex net differential effect.

Thanks to the high precision in \logg, gravity-sensitive (majority) species such as \ion{Fe}{ii} and \ion{Ti}{ii} are the least susceptible to errors in the primary stellar parameters, as shown in Sect.~\ref{sec:uncertainties}.

\begin{table*}
\caption{Photometry, data quality and stellar parameters.}
\label{Tab:SPs}
\centering
\begin{tabular}{lrc ccc cccc ccccc}
\hline\hline
 ID & \ch{RA(J2000)} & $u$ & $v$ & HR9 & HR10 & HR11 & HR15 & Teff & log g & vmic \\ 
 & \ch{Dec(J2000)} & $b$ & $y$ & (S/N) & (S/N) & (S/N) & (S/N) & (K) & (dex) & (\kms) \\ \hline
NGC6752-1003   & 19 10 50.351 &  17.423 &  15.974 &   134 & \dots & \dots & \dots &5174 & 3.10 &  1.24 \\
& $-60$ 02 44.166 &  16.575 &  15.479 \\
NGC6752-1013   & 19 10 56.817 &  17.431 &  15.969 &   126 & \dots & \dots & \dots & 5173 & 3.09 &  1.24 \\
& $-60$ 03 06.365 &  16.579 &  15.464 \\
NGC6752-1028   & 19 11 24.854 &  17.445 &  15.985 &   126 & \dots & \dots & \dots & 5174 & 3.10 &  1.24 \\
& $-59$ 57 06.584 &  16.590 &  15.475 \\
NGC6752-1112   & 19 10 35.277 &  17.585 &  16.096 &   103 &    84 &    96 & \dots & 5192 & 3.16 &  1.25 \\
& $-59$ 57 06.024 &  16.692 &  15.615 \\
\hline
\end{tabular}
\tablefoot{
 The full table can be retrieved from CDS/Vizier.
}
\end{table*}

\subsection{Spectrum synthesis}
For our heterogeneous sample of spectra of 194 stars, we opted for an automated spectrum analysis rather than the manual line-by-line approach employed for UVES spectra in Paper~I. We use a modified version of the spectrum synthesis code Spectroscopy Made Easy \citep[SME,][]{Valenti1996,Valenti2005}, set to run without user interaction. Additionally, the code has been modified to allow non-LTE (NLTE) line formation, from a grid of precomputed corrections (tabulations of LTE departure coefficients). To run the code, one provides an input spectrum, stellar parameters (\Teff, \logg, $\FeH$ and \vmic), a line list with line parameters, line and continuum masks, and a list of spectrum segments wherein the continuum is individually normalized. The code uses a grid of MARCS plane-parallel and spherically symmetric model atmospheres  \citep{Gustafsson2008}, all with scaled solar abundances and alpha-enhancement of 0.4\,dex. Interpolating models from this grid, spectra are then computed on the fly. The numerical comparison between synthetic and observed spectra is executed by a non-linear optimization algorithm \citep{Marquardt1963,Press1992}. We apply NLTE corrections for Fe, Na and Li \citep{Lind2009, Lind2011,Lind2012} to the line formation using pre-computed departure coefficients. Finally, we apply NLTE corrections to our LTE abundances of Mg (Y. Osorio, priv. comm.) as in Paper~I. \\

\subsection{Microturbulence}
As a large fraction of our stellar sample only has spectra of moderate quality in the rather line-poor spectral region covered in HR15B, \vmic\ cannot be determined to acceptable precision for each star. We executed preliminary analyses where \vmic\ was determined from a set of iron lines in the HR9B setting, and found an essentially linear dependence on effective temperature with an RMS error of 0.11\, km\,s$^{-1}$. Applying this linear relation to all stars, \vmic\ values range between 1.8\,\kms\ at \Teff\ $=6000$\,K and 1.3\,\kms\ at \Teff\ $=5200$\,K. On our photometric temperature scales, the excitation equilibrium is not necessarily fulfilled. This could skew the determination of \vmic\ due to the tendency for low-$\chi$ lines to be stronger, thus correlating \Teff\ with \vmic. We executed test runs where \Teff\ was determined simultaneously with \vmic\ from \ion{Fe}i lines, resulting in overall significantly higher temperatures, as well as a somewhat expanded temperature scale. The difference is $\delta $\Teff\ $= 207 \pm 143$\,K and $60\pm48$\,K for the warm (\Teff\ $> 5800$\,K) and cool (\Teff\ $< 5450$\,K) stars of our sample, with corresponding increases $\delta$\vmic\ $=0.10 \pm 0.16$ and $0.08\pm0.07$\,\kms, respectively. Deriving \vmic\ from \ion{Fe}{ii} lines, as was done in Paper I, we find only a minor difference $\delta$\vmic\ $=-0.06 \pm 0.20$\,\kms\ for the warm, and $\delta$\vmic\ $=0.01\pm0.06$\,\kms\ for the cool stars. With these considerations in mind, we adopt 0.10\,\kms\ as our representative uncertainty. At this level, the \vmic\ uncertainty does not add significantly to the error budget of derived abundances (cf. Sect.~\ref{sec:uncertainties}).

\subsection{Deriving chemical abundances}
We derive chemical abundances for each star independently from each available spectral order. This allows us to check the order-to-order consistency, and ensures internal homogeneity among abundances derived from a single spectral order. Only after ensuring consistent results do we average results from different spectral orders, where relevant. As spectral orders vary in the amount and quality of spectral information, they are weighted by statistical uncertainties in the averaging.

To further maintain consistency between settings and reduce (internal) line-to-line scatter, we determine abundances differentially to a high-quality averaged spectrum on the cool end of our sample for the $\alpha$ and iron-peak elements. We selected all 11 stars on the bRGB (\Teff\ $< 5420$\,K) with complete sets of observations in all spectral orders. By coadding these spectra, we achieve S/N well above 200 in all settings. We executed a preliminary abundance analysis on these spectra, and adopted these results as our reference abundance pattern. Line oscillator strengths were then adjusted to reproduce this abundance pattern, so that the baseline (at the bRGB) is the same for all lines, in all spectral orders. The reference stars are detailed in Table~\ref{Tab:SN-coadd}.

\begin{table}
\caption{Coadded bRGB spectra used as reference in differential analysis.}
\label{Tab:SN-coadd}      
\centering
\begin{tabular}{lcccccccc} 
\hline\hline
Star ID      & S/N & S/N & S/N & S/N & \Teff & \logg \\ 
      		& HR9 & HR10 & HR11 & HR15 & (K) & (dex) \\ \hline
1872         &  89 &  72 &  64 &  63&  5302 & 3.46 \\
1980         &  76 &  69 &  90 & 132&  5309 & 3.48 \\
2113         &  73 &  63 &  63 &  58&  5328 & 3.51 \\
2089         &  80 &  65 &  68 &  57&  5328 & 3.51 \\
2129         &  73 &  69 &  70 &  59&  5332 & 3.52 \\
2176         &  71 &  61 &  97 & 135&  5339 & 3.54 \\
201217     &  56 &  55 &  80 & 114&  5355 & 3.57 \\
2334         &  69 &  58 &  74 & 106&  5355 & 3.57 \\
2396         &  76 &  57 &  87 & 117&  5358 & 3.57 \\
2337         &  68 &  61 &  90 & 123&  5361 & 3.57 \\
2646         &  75 &  67 &  86 & 106&  5413 & 3.62 \\\hline
coadded   & 224 & 212 & 267 & 259 & 5344 & 3.54 \\ 
  \hline\hline
\end{tabular}
\end{table}

When the chemical composition exhibits internal variations among the coadded stars, nontrivial problems may arise from the averaging of lines whose strengths and shapes are intrinsically different. For this reason, we do not adopt astrophysical oscillator strengths for the light elements where such variations are expected: Li, Na, Mg, and Al. Note however that most of these elemental abundances are derived from only one or two lines, and thus would anyway not be affected by this issue. The atomic data for these lines can be found in Table~\ref{Tab:atomic data}. The other atomic data for the lines used in the abundance analysis were obtained from the Vienna Atomic Line Database \citep{piskunov1995,Kupka1999}.

\begin{table}
\caption{Atomic Line Data.}
\label{Tab:atomic data}
\centering          
\include{linelist}
\end{table}

\subsection{Uncertainties} \label{sec:uncertainties}
We consider systematic and statistical errors separately.
Leading systematic error sources likely stem from uncertainties in the atmospheric parameters, which were discussed in Sect.~\ref{subsect:phot} and \ref{subsect:grav}. Systematic abundance uncertainties from the influence of stellar parameter perturbations on average abundance results are indicated in Table~\ref{Tab:abund sensitivity}. We evaluate these effects for the hot and cool ends of our temperature range, for all chemical elements analysed in this work.

Modelling errors related to the use of 1D atmospheres were discussed in Paper~I.
Lines of the same ionisation stage respond, in general, in a similar way to changes in atmospheric parameters. Thus the systematic uncertainties related to the atmospheric parameters partially cancel when working with abundance ratios ($\XY = [{\rm X}/{\rm H}] - [{\rm Y}/{\rm H}]$).

For statistical errors, we use those reported by the $\chi^2$ minimization algorithm of SME \citep{Marquardt1963,Press1992}, which are based on photon-noise statistics. We shall refer to these as 'SME errors'. As they do not fully take into account uncertainties in the continuum determination, we enforce a representative minimum error of 0.02\,dex. Unless otherwise stated, uncertainties in this work refer to these statistical errors.

\begin{figure}
\includegraphics[]{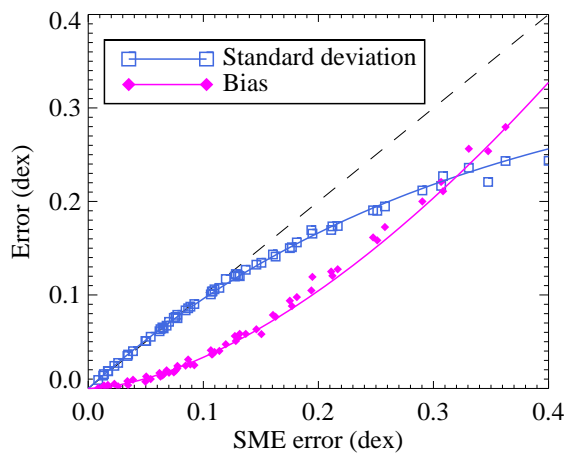}
\caption{Weighted standard deviations and biases of the weighted means as determined from weak lines analysed in Monte Carlo tests on synthetic spectra.} 
\label{fig:errors}
\end{figure}

We examine the estimate of these statistical errors in more detail by running a Monte Carlo simulation, in which we derive the abundance from a single weak atomic line, representing the case of lithium. As reference, we use synthetic spectra generated for stellar parameters (\Teff\ $=5400$--6000\,K) and abundances spanning the ranges of the stars observed ($EW = 7$--45\,m\AA), with a wide range of Gaussian noise injected ($S/N = 10$--1000). We generated a total of 65 such sets of at least 10\,000 spectra each. We averaged results from each set of abundance analyses, taking SME errors as uncertainties. The weighted standard deviation and the bias of the mean from these averages are compared in Fig.~\ref{fig:errors} to the SME error. Small errors appear to be dominated by the width of the distribution, such that averaging a large sample will generate an accurate estimate of the mean. Large errors on the other hand tend to become asymmetric, giving a significant positive bias, while underestimating the true dispersion. The figure can be interpreted as a decomposition of the SME error into these two components, for any single abundance result. Or, the bias of an averaged abundance can be estimated from the size of its standard deviation. For instance, averaging a large sample of lithium abundances with 0.10\,dex uncertainties would overestimate the mean value by 0.03\,dex.

The cause of this behaviour is that noise produces errors in line equivalent widths which are symmetric and linear. For weak lines, the same is then true for the number of absorbers $N_{\rm X}/N_{\rm H}$. On the logarithmic abundance scale, $(\log N_{\rm X}/N_{\rm H})$, these errors instead appear asymmetric, such that weighted averages always systematically overestimate the mean value. If instead lines are on the saturated or strong parts of the curve of growth, the corresponding error statistics apply. Finally, these considerations would not necessarily apply if uncertainties in stellar parameters dominate over those in line strengths.

As we analyze lines on various parts of the curve of growth, we choose not to apply linear averages. Tests indicate that the magnitude of these biases, if applicable, is less than 0.03\,dex in this work.

\begin{table}
\caption{Abundance sensitivity to stellar parameters.}
\label{Tab:abund sensitivity}
\centering
\begin{tabular}{l r r r r r r}\hline\hline \noalign{\smallskip}
Ion &  \multicolumn 2 c {$T_{\mbox{{\scriptsize eff}}} + 100$\,K}  & \multicolumn 2 c {$\log g + 0.1$\,dex} & \multicolumn 2 c {$v_{\mbox{{\scriptsize mic}}} + 0.1$\,\kms} \\ 
            & \ch{TOP}    &   \ch{bRGB}    &  \ch{TOP}   &   \ch{bRGB}   &   \ch{TOP}   &    \ch{bRGB} \\  \hline\noalign{\smallskip}
Li I   & $ 0.073$ & $ 0.078$ & $-0.002$ & $-0.002$ & $-0.001$ & $ 0.006$  \\
Na I   & $ 0.032$ & $ 0.046$ & $ 0.002$ & $ 0.000$ & $ 0.000$ & $-0.001$  \\
Mg I   & $ 0.080$ & $ 0.116$ & $-0.030$ & $-0.050$ & $-0.014$ & $-0.007$  \\
Al I   & \dots\hphantom{..} & $ 0.037$ & \dots\hphantom{..} & $ 0.001$ & \dots\hphantom{..} & $ 0.004$  \\
Si I   & $ 0.025$ & $ 0.036$ & $ 0.004$ & $ 0.008$ & $ 0.009$ & $-0.001$  \\
Ca I   & $ 0.048$ & $ 0.065$ & $ 0.000$ & $-0.002$ & $-0.007$ & $-0.015$  \\
Sc II  & \dots\hphantom{..} & $ 0.043$ & \dots\hphantom{..} & $ 0.035$ & \dots\hphantom{..} & $-0.006$  \\
Ti II  & $ 0.038$ & $ 0.050$ & $ 0.032$ & $ 0.026$ & $-0.010$ & $-0.025$  \\
Cr I   & \dots\hphantom{..} & $ 0.132$ & \dots\hphantom{..} & $-0.020$ & \dots\hphantom{..} & $-0.038$  \\
Fe I   & $ 0.075$ & $ 0.109$ & $-0.006$ & $-0.014$ & $-0.014$ & $-0.027$  \\
Fe II  & $ 0.040$ & $ 0.058$ & $ 0.022$ & $ 0.013$ & $-0.019$ & $-0.031$  \\
Ni I   & \dots\hphantom{..} & $ 0.073$ & \dots\hphantom{..} & $ 0.004$ & \dots\hphantom{..} & $-0.004$  \\
 \hline\hline
\end{tabular}
\tablefoot{Effects on the averaged abundances are shown for the hot and cool ends of our sample, as defined in Table~\ref{Tab:mean-trends}.
}
\end{table}

 %#######################################################################################

%******************************************************************
%                                   RESULTS
%******************************************************************

\section{Results}
In what follows we address our derived chemical abundances for the total sample of 194 stars. As not all stars were observed with all settings, we do not have a complete set of abundances for the full sample. Results from different spectral orders have been averaged only after carefully verifying that they appear compatible. After describing our abundance results, we discuss the implications, and investigate the prospect of predicting population membership from photometry.

\subsection{Chemical abundances}
\subsubsection{Lithium} \label{sec:lithium}
Lithium abundances $A$(Li) are given in Table~\ref{Tab:pop-abun} and are derived from the \element[][7]{Li} resonance line at 6707.8\,\AA. The line has two fine-structure components which are separated by only 0.15\,\AA\ and thus are indistinguishable at the resolution of GIRAFFE ($R = 19\,300$). The atomic data for the Li doublet takes into account this fine structure and isotopic splitting and is given in Table~\ref{Tab:atomic data}. The abundances are corrected for NLTE effects by interpolation on the grid by \citet{Lind2009}. As lithium is almost completely ionised, $A$(Li) is primarily sensitive to \Teff, which thus dominates the systematic errors. Our estimated precision of 40\,K in \Teff\ corresponds to an abundance uncertainty of 0.03~dex. Compared to this, errors due to uncertainties in gravity, metallicity and microturbulence are generally negligible ($\sim 0.01$\,dex).

Combining these systematic uncertainties, with the expected bias stemming from the averaging process, we estimate 0.10\,dex to be a suitable conservative error. The average Li abundance of the TOP sample (\Teff > 5900\,K) is $2.22\pm0.08$ in agreement with Paper~I and \citet{Gratton2001}. We will discuss lithium in greater depth in Sect.~\ref{sec:evol-lithium}.

\subsubsection{$\alpha$ and iron-peak elements} \label{sec:aip}
Abundances of the $\alpha$ elements magnesium, calcium and titanium were derived along with the iron-peak elements scandium, chromium, iron and nickel.
\begin{figure*}
\begin{center}
\includegraphics[]{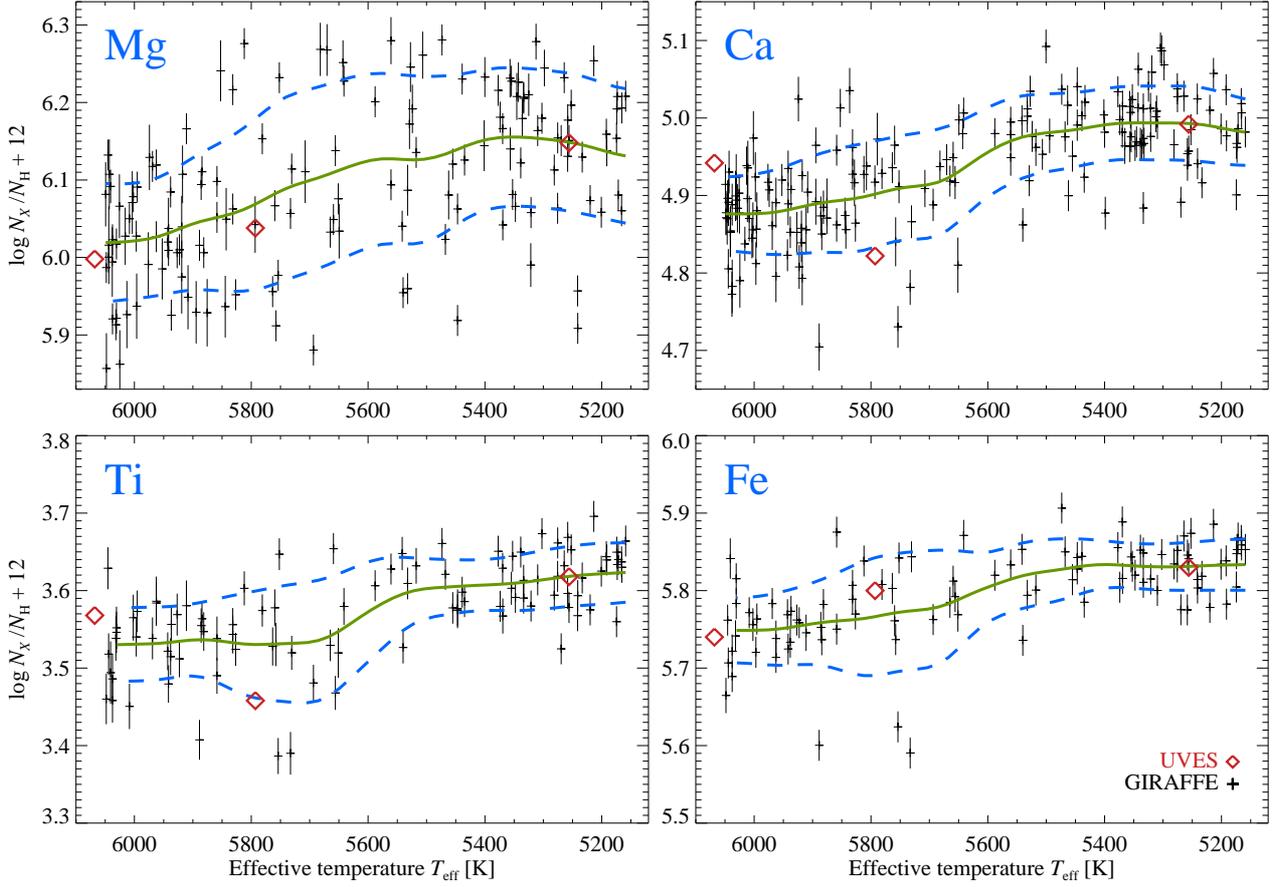}
\caption{Evolutionary abundance trends of Mg, Ca, Ti, and Fe. Mg and Ca abundances are derived from neutral lines while Ti and Fe are derived from singly ionised lines (see Table~\ref{Tab:mean-trends}). The solid (green) line represents the running mean (weighted average), while dashed (blue) lines represent the standard deviation. Error bars represent statistical errors. Systematic errors are indicated in Fig.~\ref{Fig:AD-trends}.
Overplotted in diamonds are the UVES results from Paper~I, which have been arbitrarily shifted for easier comparison.}
\label{Fig:abundances}
\end{center}
\end{figure*}
Figure~\ref{Fig:abundances} shows the derived chemical abundances of magnesium, calcium, titanium and iron as a function of the effective temperature. Results have been averaged on a running mean in bins of $\pm$150\,K, which has then been further smoothed by a Gaussian ($\sigma = 100$\,K). 
We show also the group-averaged abundances derived from UVES spectra in Paper~I, from a temperature scale based on the $(b-y)$ colour index. 
We have shifted those results slightly to coincide with our abundance scale for the coolest stars.

Magnesium abundances were derived from the $\lambda\lambda$5167--5184 triplet in the HR9 setting and the high-excitation $\lambda$5528 line in the HR10 setting. 
We average abundances between the two settings for the 50 stars which were observed in both settings, as we find rather consistent results when adopting laboratory oscillator strengths \citep{Chang1990,Aldenius2007}  and applying NLTE corrections. The rather large observed dispersion, compared to the other elements, is further discussed in Sect.~\ref{sec:anticorr}.

Calcium abundances were likewise determined from the HR9, HR10 and HR11 settings. When applicable, we average results between settings, resulting in abundances determined for a set of 176 stars. We investigated departures from LTE, following \citet{Mashonkina2007}, for several of the \ion{Ca} i lines used in the determination of the Ca abundance. Apart from the general effect of over-ionisation which raises all abundances by $\sim 0.1$\,dex, the relative trends between TOP and RGB are only marginally affected, on average by 0.02\,dex. This correction is well within the error budget of our analysis. 

Titanium and iron abundances have been derived from ionized lines, which we selected exclusively from HR9. As only half the stars were observed with this setting, we also determined iron abundances for each star from the neutral lines available in every spectral order. These abundances were corrected for NLTE effects and averaged between settings, but are only used as a reference in computing \XFe.

\begin{table*}
\caption{Average abundances obtained at two effective temperature points.}
\label{Tab:mean-trends}
\centering
\begin{tabular}{lcccrrrr}\hline\hline
Group & $T_{\mbox{\scriptsize eff}}$ & $\log g$ & $\xi$ & \ch{$\abund{Mg}$\tablefootmark{a}} & \ch{$\abund{Ca}$\tablefootmark{b}}  & \ch{$\abund{Ti}$\tablefootmark{c}}  & \ch{$\abund{Fe}$\tablefootmark{d}} \\ 
	      & (K)     & (cgs) & (km\,s$^{-1}$) & \ch{NLTE} & \ch{LTE} & \ch{LTE} & \ch{NLTE}  \\ \hline
TOP       & 6050 & 4.02 & 1.8 & $ 6.02 \pm0.08$ & $ 4.88 \pm0.05$ & $ 3.53 \pm0.05$ & $ 5.75 \pm0.04$ \\
bRGB      & 5160 & 3.03 & 1.2 & $ 6.13 \pm0.09$ & $ 4.98 \pm0.04$ & $ 3.62 \pm0.04$ & $ 5.83 \pm0.03$ \\
$\Delta(\text{TOP}-\text{bRGB})$  & 890 & 0.99 & 0.6 & $-0.11 \pm0.12$ & $-0.11 \pm0.06$ & $-0.09 \pm0.06$ & $-0.08 \pm0.05$ \\
 \hline\hline
\end{tabular}
\tablefoot{
\tablefoottext{a}{Based on Mg\,{\scriptsize I} $\lambda\lambda$5167, 5173, 5184 and 5528.}
\tablefoottext{b}{Based on Ca\,{\scriptsize I} $\lambda\lambda$5261, 5262, 5265, 5349, 5582, 5589, 5590, 5594, 5598, 5601 and 5603.}
\tablefoottext{c}{Based on Ti\,{\scriptsize II} $\lambda\lambda$5154, 5185, 5188, 5226 and 5336.}
\tablefoottext{d}{Based on Fe\,{\scriptsize II} lines $\lambda\lambda$5146, 5169, 5197, 5234, 5264, 5276, 5284 and 5316.}}
\end{table*}

All elements exhibit clear evolutionary variations, summarized for the hot and cool ends of our temperature range in Table~\ref{Tab:mean-trends}, where we also state explicitly which spectral lines have been used. 
Although these abundance variations are rather weak and of low significance (1--$2\sigma$), they are in good agreement with Paper~I, despite the independent methods used here. The abundance variations are also rather robust regarding systematic errors in stellar parameters. For instance, removing the abundance trend in iron would require increasing temperatures of stars at the TOP by 200\,K, implying an error in $\Delta\text{\Teff}(\text{TOP}-\text{bRGB})$ of 25\,\%, which we consider unlikely. A similarly large adjustment to \logg\, (+0.35\,dex) or \vmic\ (--0.4\,\kms) would be required. Systematic uncertainties of stellar parameters were further examined by \citet{Nordlander2012}, indicating that no combination of stellar parameters can simultaneously render null trends in all elements. Errors due to hydrodynamical effects were investigated in Paper~I. A full 3D treatment (in LTE) would in fact strengthen all observed trends.

\begin{table}
\caption{Average abundances and dispersions of Sc, Cr, Ni and Fe for the subsample of bRGB stars (\Teff\ $<5450$\,K).}
\label{Tab:abundances}
\centering
\begin{tabular}{lcccc}\hline\hline
 & Sc\tablefootmark{a}  & Cr\tablefootmark{b} & Fe\tablefootmark{c} & Ni\tablefootmark{d}\\ 
 & LTE & LTE & NLTE & LTE \\ \hline
$\log\varepsilon$ & 1.52 & 3.88 & 5.83 & 4.48 \\
$\sigma(\log\varepsilon)$  & 0.05 & 0.05 & 0.03 & 0.03 \\
  \hline\hline
\end{tabular}
\tablefoot{The dispersion refers to the weighted standard deviation around the mean.
\tablefoottext{a}{Based on Sc\,{\scriptsize II} $\lambda\lambda$5527 and 5658.}
\tablefoottext{b}{Based on Cr\,{\scriptsize I} $\lambda\lambda$5206--5208.}
\tablefoottext{c}{Based on Fe\,{\scriptsize II} lines $\lambda\lambda$5146, 5169, 5197, 5234, 5264, 5276, 5284 and  5316.}
\tablefoottext{d}{Based on Ni\,{\scriptsize I} $\lambda\lambda$5146, 5155, 5477, 6643 and 6768.}
}
\end{table}

We also derived abundances of scandium, chromium and nickel for the bRGB stars in our sample. Unfortunately, the quality of the dwarf star spectra did not allow to derive reliable abundances for these elements in these stars. An overview of the mean abundances and dispersion for the bRGB stars (\Teff\ $< 5450$\,K) is given in Table~\ref{Tab:abundances}. \\

The abundances of Ca, Ti, Cr, Fe and Ni are in agreement with the abundances derived for the reference RGB bump star NGC6752--11 from the \citet{Yong2013} sample. Unfortunately, they did not derive Mg or Sc abundances. The abundances show the typical pattern of halo field stars in which the $\alpha$-elements are overabundant by 0.1 to 0.3\,dex relative to the Fe-peak elements which are roughly solar, meaning $\left[\alpha\mbox{/Fe}\right]\approx0.1$--0.3 and $\left[\mbox{X/Fe}\right]\approx0$. \\

Three of the stars examined here (IDs 2844, 2795, 3035) exhibit suspiciously low abundances in calcium, titanium and iron, which could indicate binarity. For instance, an evolved low-mass companion would only moderately affect photometric parameters, but rather provide continuum flux filling in all spectral lines. 

\begin{figure*}
\begin{center}
\includegraphics[]{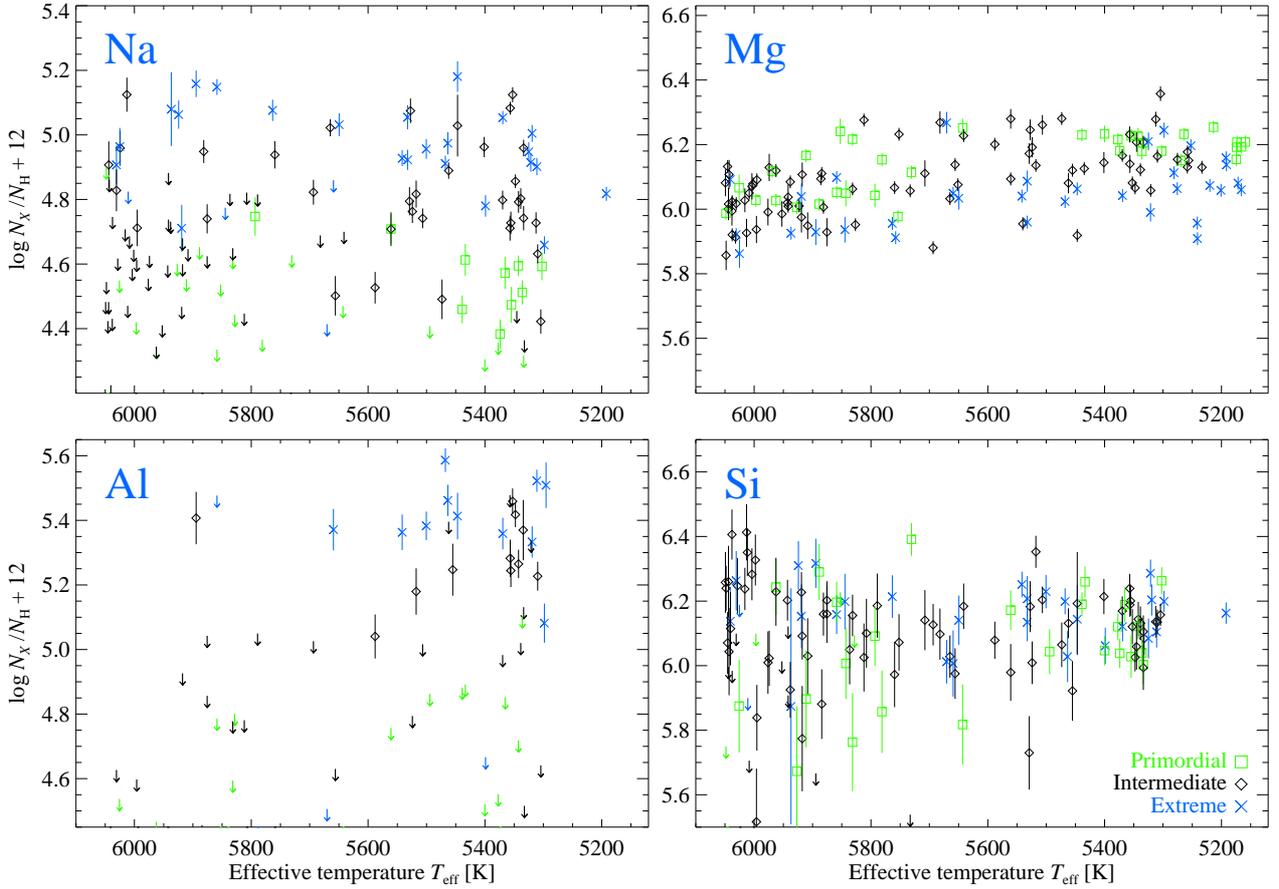}
\caption{Evolutionary abundance trends of Na, Mg, Al, and Si. The different coloured symbols correspond to the chemical populations deduced from a photometric cluster analysis (see text).}\label{Fig:pop_abun}
\end{center}
\end{figure*}

\subsubsection{Na, Mg, Al and Si} \label{sec:anticorr}
Distinct chemical populations in globular clusters have previously been addressed in a series of papers (on NGC\,6752, e.g., by \citealt{Gratton2001} and \citealt{Carretta2007a, Carretta2012}).
In Paper~I, we did the same for a set of FLAMES-UVES spectra, revealing three chemically distinct populations in agreement with \citet{Carretta2012}. Based on the abundances of Al, Na and Mg we labelled stars to be first- or second-generation stars. First-generation stars are poor in Al and Na while rich in Mg, compared to the mean of the whole sample. This group of stars is referred to as the primordial population (P). Second-generation stars are characterised by average and high Al and Na abundances compared to the mean abundances for our sample. This second generation can be further split into two different populations: one with average abundances of Al, Na and Mg (intermediate population I), and one rich in Al and Na but poor in Mg (extreme population E). Si is not considered a discriminating element in this sense. \\

Figure~\ref{Fig:pop_abun} shows the abundances of sodium, magnesium, aluminium and silicon as a function of \Teff. All elements exhibit a large dispersion, as expected from previous findings (e.g. \citealt{Carretta2007a,Carretta2012}; Paper~I). 
Symbols and colours in this figure correspond to the different chemical populations, determined photometrically in Sect.~\ref{sec:clusteranalysis}.

Sodium abundances were derived for a set of 132 stars using the $\lambda\lambda$5682--5688 doublet in HR11. We find a spread (tip-to-tip) of roughly 0.8\,dex, which does not appear to vary significantly along the sequence, sans sampling effects. While \citet{Lind2011} found a 0.2\,dex difference between dwarf and giant stars in the cluster NGC\,6397, we do not observe a clear difference in abundance between the two groups. \citet{Lind2011} explain their observed difference as a result from atomic diffusion. At the present metallicity, the expected trend is shallower ($\sim 0.1$\,dex), while the star-to-star dispersion is similar. A potential trend is thus difficult to detect using the very weak lines available in our spectra. We will address atomic diffusion in more detail in Sect.~\ref{subsect:AD}.

By contrast, our magnesium abundances exhibit a smaller spread of just 0.3\,dex. The spread again does not exhibit significant variation along the sequence. The abundance difference comparing warm to cool stars, $0.11\pm0.12$\,dex (Table~\ref{Tab:mean-trends}), is similar for the three populations, $0.16 \pm 0.06$, $0.16 \pm 0.12$ and $0.12 \pm 0.14$\,dex, for the P, I and E populations, respectively. As the P and I populations exhibit similar abundances of magnesium, we note that their combined sample indicates an abundance trend of $0.15 \pm 0.08$\,dex.

Aluminium abundances are derived from the weak \ion{Al} i doublet at 6696--6698\,\AA\ and thus could only be derived from stars observed in HR15. Unfortunately, the doublet was too weak to be observed in most of the stars, and even upper limits were not meaningful for the bulk of (lower-S/N) spectra. 
In total, we could detect aluminium in 22 stars, and derived upper limits for another 48, indicating a spread of at least 1\,dex.

Finally, we deduced silicon abundances from three Si\,{\scriptsize I} lines in HR11B at 5708, 5665 and 5690\,\AA. The spread in Si abundance is dominated by the observational uncertainty on the hotter end of our sample. This uncertainty also masks out any potential trend with evolutionary phase. We estimate an intrinsic spread in Si of about 0.3\,dex.

\begin{figure*}
\begin{center}
\includegraphics[]{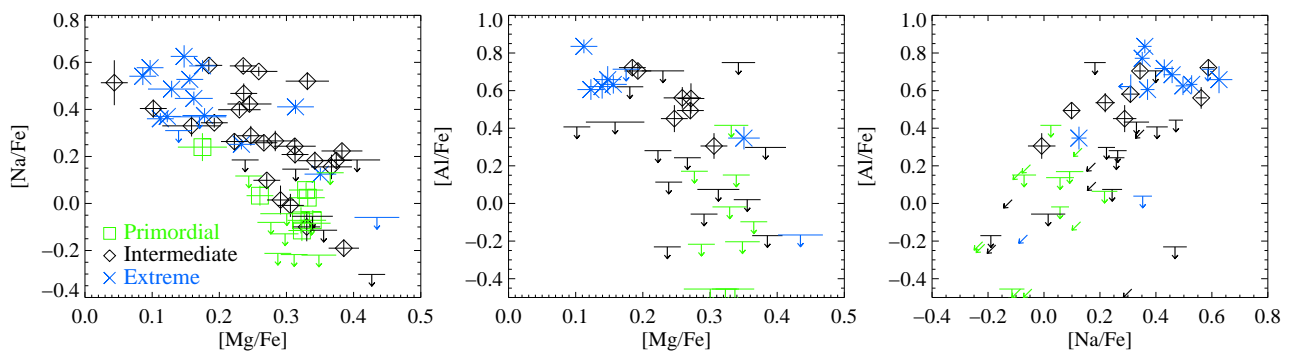}
\caption{Anticorrelations between sodium, magnesium and aluminium. Different coloured symbols correspond to the chemical populations deduced from a photometric cluster analysis (see text). The faintest ($V > 16.8$) stars have been excluded from this comparison.}\label{Fig:AbunRatio}
\end{center}
\end{figure*}

The abundances ratios of Na, Mg, Al and Si relative to Fe (based on \ion{Fe} i lines) for the individual stars are given in Table~\ref{Tab:pop-abun}, and compared in Fig.~\ref{Fig:AbunRatio}. Using Mg as the discriminator, the usual anticorrelations are observed, a signature of element depletion in H burning at high temperature. We also see that Al is well correlated with elements which are enhanced by the Ne-Na and Mg-Al cycles. And like \citet{Carretta2012}, we also seem to find a Al-Si correlation. The correlation stems from leakage from the Mg-Al cycle on $^{28}$Si. Such a leakage can only be obtained within a reaction network in which the temperature exceeds $\sim 65 \times 10^6$\,K. The Al-Si correlation can thus be explained as a direct result of pollution by the first generation of cluster stars burning H at high temperatures.

\begin{table*}
\caption{Light element abundances and abundance ratios, compared to photometric population assignments.}
\label{Tab:pop-abun}      
\centering          
\begin{tabular}{lccccccc} 
\hline\hline
      Star      ID	   & $A$(Li) & \NaFe  &  \MgFe 	& \AlFe & \SiFe &  $\log\varepsilon$(Fe) & Pop\tablefootmark{a}  \\ 
                       & NLTE    & NLTE   &  NLTE   & LTE   & LTE   &  NLTE                  &                       \\ \hline
1802    & $ <0.82$ & \dots        & \dots        & $   0.78 \pm   0.08$ & \dots        & $   5.81 \pm   0.03$ & E \\
1980    & $   1.31 \pm   0.04$ & $   0.10 \pm   0.04$ &  $   0.27 \pm   0.03$ & $   0.49 \pm   0.05$ & $   0.26 \pm   0.04$ & $   5.81 \pm   0.02$ &      I \\
1983    & $ <1.05 $ & $   0.50 \pm   0.03$ & \dots        & $   0.63 \pm   0.05$ & $   0.36 \pm   0.05$ & $   5.79 \pm   0.02$ &      E \\
2117    & $   1.19 \pm   0.07$ & $   0.03 \pm   0.04$ & $   0.33 \pm   0.03$ & $ <0.42 $ & $   0.21 \pm   0.05$ & $   5.76 \pm   0.02$ &      P \\
2129    & $   1.34 \pm   0.09$ & $   0.27 \pm   0.04$ & $   0.22 \pm   0.03$ & $ <0.29 $ & $   0.25 \pm   0.06$ & $   5.81 \pm   0.02$ &      I \\
2151    & $   1.00 \pm   0.07$ & $   0.31 \pm   0.04$ & \dots        & $   0.58 \pm   0.05$ & $   0.32 \pm   0.06$ & $   5.76 \pm   0.02$ &      I \\
\hline\hline
\end{tabular}
\tablefoot{Uncertainties are given by statistical errors. All abundances are based on lines of the neutral species. Upper limits are given by $<$.\\
\tablefoottext{a}{Photometrically identified chemical population: primordial (P), intermediate (I) and extreme (E).}\\ The full table can be retrieved from CDS/Vizier. \\
Solar reference abundances: $A(\text{Na}) = 6.17$, $A(\text{Mg}) = 7.53$, $A(\text{Al}) = 6.37$, $A(\text{Si}) = 7.51$, $A(\text{Fe}) = 7.45$.
}
\end{table*}

In the next section we will address these distinct chemical populations in greater depth by investigating their photometric properties.

\subsection{A cluster analysis}\label{sec:clusteranalysis}
It has previously been demonstrated by \citet{Carretta2012} that one can couple abundance anticorrelations to photometric properties of RGB stars in NGC\,6752. We shall here attempt to extend their findings for fainter stars, along the SGB toward the TOP region. 

\begin{figure}
\begin{center}
\includegraphics[]{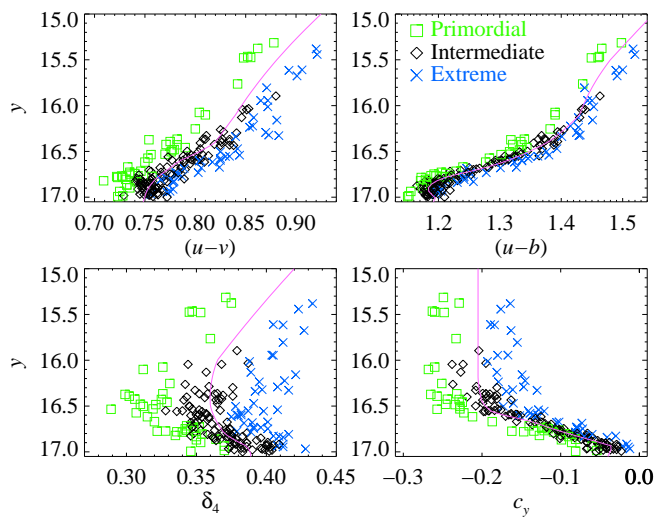}
\caption{CMDs in $(u-v)$ (upper left), $(u-b)$ (upper right), $\delta_4$ (lower left), and $c_y$ (lower right). The solid lines represent the fiducial sequences from which distances (residuals) are used in the cluster analysis. Different coloured symbols correspond to the chemical populations deduced from a cluster analysis.}\label{Fig:CMD-abun}
\end{center}
\end{figure}

Using $uvby$ Str\"omgren photometry for our complete data set, we created colour-magnitude fiducial sequences for the colour indices $(u-v)$, $(u-b)$, $c_y$ and $\delta_4$. Figure~\ref{Fig:CMD-abun} shows the corresponding CMDs with the fiducial sequences overplotted. The $c_y$ index was defined by \citet{Yong2008} and combines three colour indices: $c_y = (u-v) - (v-b) - (b-y) = u - 2\times v + y$. It is sensitive to N abundances, from the influence on NH bands near 3450\,\AA\ in the $u$ filter, and CN bands at 4216\,\AA\ in the $v$ filter. As the sensitivity on the latter is weaker, it is weighted twice. The $\delta_4 = (u-v) - (b-y)$ index was introduced by \citet{Carretta2011} and combines the four Str\"omgren filters, each weighted only once. For further in-depth discussion of the $c_y$ and $\delta_4$ indices, the reader is referred to \citet{Carretta2011}. \\

We now pose the following question: Based solely on photometry, can we predict membership of our globular cluster stars to different chemical populations? It was shown by \citet{Sbordone2011} and \citet{Cassisi2013} that the effect of anticorrelated Mg and Al variations on Johnson and Str\"omgren photometry is negligible. However, these anticorrelations appear intrinsically tied to variations in CNO and He, which affect the spectral energy distribution (SED) both directly via molecular band strengths, and indirectly via secondary effects on the stellar evolution. We here make an attempt to predict chemical population membership based on photometric information alone. In a next step we then compare these predictions to the derived abundances for Na, Mg, Al and Si. This was done as follows.

With the photometric indices described above, we performed a cluster analysis using the $k$-means algorithm \citep{Steinhaus1956,MacQueen1967} as implemented in the $R$ statistical package \citep{R2013}\footnote{http://www.R-project.org}. We consider only photometric parameters for each star and take the residuals of colour indices with respect to the fiducial sequences at constant $V$ magnitude as the parameters for the cluster analysis. We do not take into account the varying widths of the distributions as a normalizing factor. By the nature of the $k$-means method, uncertainties are not taken into account, and populations are assigned as definitive rather than probabilistic. The resulting average values of the photometric parameters for stars in three populations as determined by our cluster analysis are given in Table~\ref{Tab:clusters}. \\

\begin{table}
\caption{Average photometric parameters of cluster populations}
\label{Tab:clusters}      
\centering          
\begin{tabular}{lccccc} 
  \hline\hline
 Pop. & nr & $\Delta u-v$ & $\Delta u-b$ & $\Delta c_y$ & $\Delta \delta_4$  \\ \hline
 P	 & 46 & -0.030 & -0.027 & -0.033 & -0.033 \\
 I	 & 101 & 0.001 & 0.001 & -0.001 & 0.001 \\
 E	 & 47 & 0.026 & 0.019 & 0.031 & 0.035 \\
  \hline\hline
\end{tabular}
\tablefoot{Stars have been identified by a $k$-means analysis on photometric data.}
\end{table}

With all stars in our sample assigned to populations, we return to the light element abundances. In Fig.~\ref{Fig:AbunRatio} we show the relations between abundances of sodium, magnesium and aluminium, colour-coded by population. We find stars low in sodium and aluminium but rich in magnesium to be generally correctly identified as first-generation stars. Stars rich in sodium and aluminium but low in magnesium are labeled as extreme stars belonging to the second generation. There is, however, some crosstalk between the primordial and intermediate population on the one hand, and between the extreme and intermediate populations on the other. 
Since many of our abundance results for these elements are only upper limits, it is difficult to say definitively whether stars have truly been mislabeled. From Fig.~\ref{Fig:CMD-abun}, more crosstalk is seen among the faintest ($V > 16.8$), warmest ($T_\text{eff} > 5900$\,K) stars of the sample. This should be expected due to weaker correlation between photometry and chemistry on the merit of weaker CN-bands. Spectroscopic assignments are unfortunately similarly difficult as a result of limited $S/N$ for these faint stars, making it difficult to derive precise abundances from weak lines (see Fig.~2 in Paper~I).

Comparing the statistics of our population assignments to those by \citet{Carretta2012}, we find overall similar results.
The primordial population comes out the smallest at 24\,\%, at a similar fraction to the 21\,\% of \citet{Carretta2012} and 25\,\% of \citet{Milone2013}. A general fraction of $\sim 30$\,\% primordial stars in GCs was determined by \citet{Carretta2013} by comparing the Na abundances of field stars with GC stars.

We find a twice larger intermediate than extreme population, while \citet{Carretta2012} find the two to be roughly comparable, and \citet{Milone2013} find a 3:2 ratio. Our result may be somewhat inflated due to shortcomings in our analysis relating to the weaker photometric sensitivity to chemical composition toward the TOP. Firstly, the photometric variations are simply weaker in relation to the observational uncertainty. Secondly, the fact that the variations weaken toward higher temperature leads to any warm star regardless of chemical composition appearing more similar to an intermediate (cool) star. 

We have also directly compared to \citet{Carretta2012} by including their stars in the cluster analysis. Crossmatching our photometric catalogue to their sample of stars, the three populations are identified in the same proportions, to within 1\,\%. Among the stars they assigned to the P, I and E populations, we misidentify 20, 31 and 17\,\%, respectively. Most mismatches are between the I and E populations, while only one star in the E population is assigned P membership.\\

We conclude that a cluster analysis based solely on $uvby$ Str\"omgren photometry can indeed separate GC stars into first- and second-generation populations on the subgiant branch, with some success even down toward the TOP. We explore this possibility further in Sect.~\ref{sec:evol-lithium}. As expected though, significant crosstalk arises when the photometric response to chemical composition weakens to a level comparable to photometric uncertainties. Finally, we would like to note that the separation of GC stars in populations based on $uvby$ Str\"omgren photometry is the direct result of abundance variations in C, N and O, and in He. However, since O correlates with Mg but anticorrelates with N, stars rich in O/Mg will be poor in N and vice versa. This variation in N abundance is then responsible for the separation in colour index upon which we base our method of identifying chemical populations. Besides the effect of varying N abundances in the different populations, \citet{Milone2013} also discussed the influence of different He fractions, $Y$. However, as we have no direct information on the He content of our stars, we can only refer to their results that the first generation of stars is characterised by a primordial He abundance $(Y\sim0.246)$, is rich in C/O/Mg and poor in N/Na/Al. The extreme population is made up of second-generation stars enhanced in He $(\Delta Y\sim0.03)$ and N/Na/Al but depleted in C/O/Mg. Finally the intermediate population also represents second-generation stars, but with intermediate He $(\Delta Y\sim0.01)$, C/O/Mg and N/Na/Al abundances.

\begin{figure}
\begin{center}
\includegraphics[]{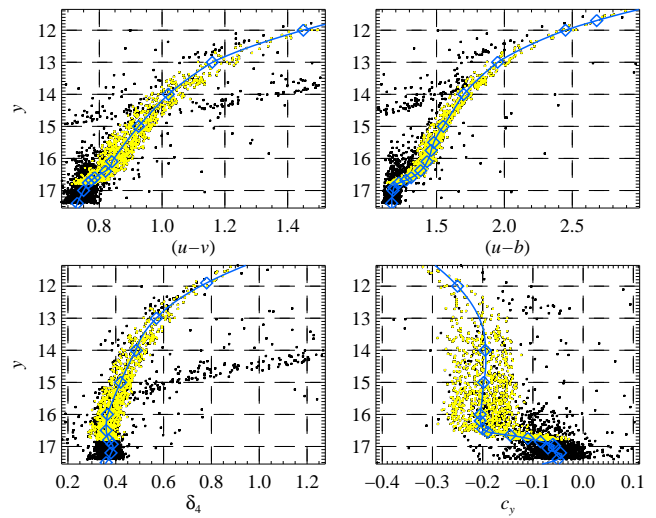}
\caption{CMDs in $(u-v)$ (upper left), $(u-b)$ (upper right), $\delta_4$ (lower left), and $c_y$ (lower right). The solid (blue) lines represent the fiducial sequences from which distances (residuals) are used in the cluster analysis. 
Blue diamonds indicate points between which the sequence has been interpolated.
Yellow squares represent the selected stars for the cluster analysis.}\label{Fig:phot selection}
\end{center}
\end{figure}

\begin{figure*}
\begin{center}
\includegraphics[]{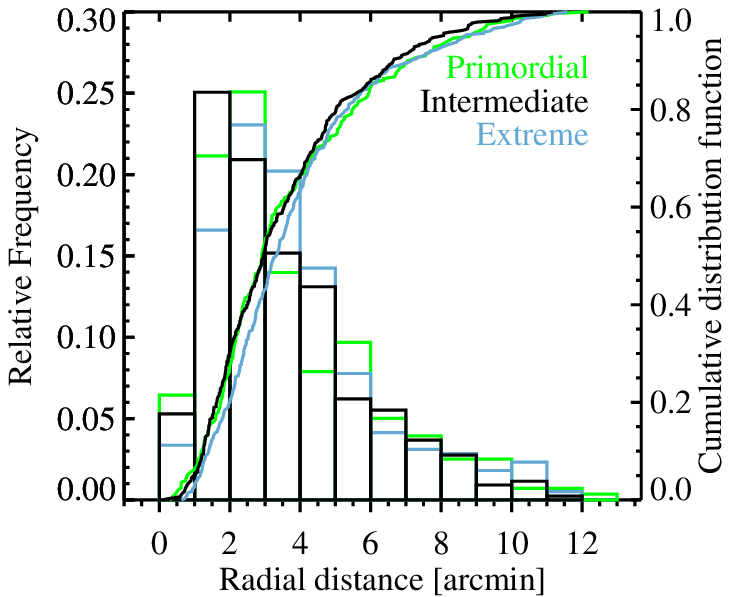}
\includegraphics[]{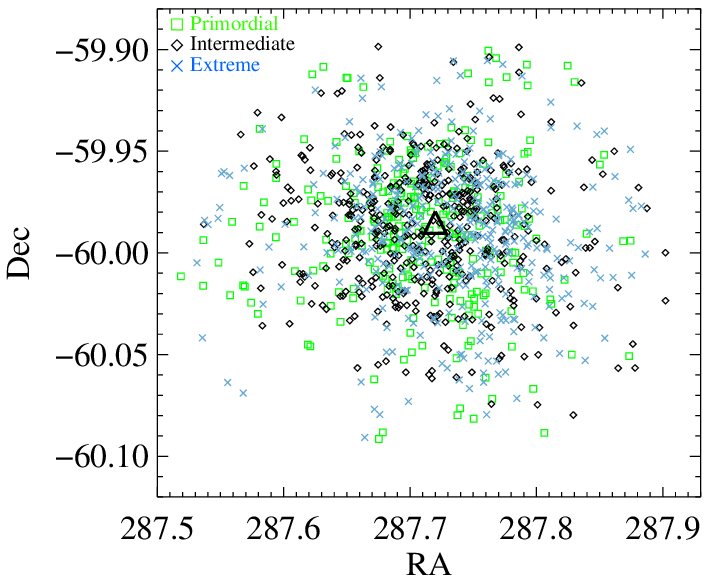}
\caption{{\sl Left}: Cumulative radial distribution of the different populations as a function of distance from the cluster center, {\sl Right}: Spatial distribution of the stars in NGC\,6752. The different symbols represent the populations as deduced from the cluster analysis. The black triangle marks the cluster center.}\label{Fig:phot cumul}
\end{center}
\end{figure*}

\subsection{Radial distribution}
Finally, we also performed the same cluster analysis on the full $uvby$ photometric catalogue after selecting all stars with $V$ magnitude brighter than 16.8 and located within a distance of 0.1\,mag from the chosen fiducials in all four colours. These criteria select 1100 stars, as shown in Fig.~\ref{Fig:phot selection}. The cumulative radial distribution for each population is given in the left panel of Fig.~\ref{Fig:phot cumul}, while the right panel displays the spatial distribution on the sky, with different symbols corresponding to the different populations. 

\begin{figure}
\begin{center}
\includegraphics[]{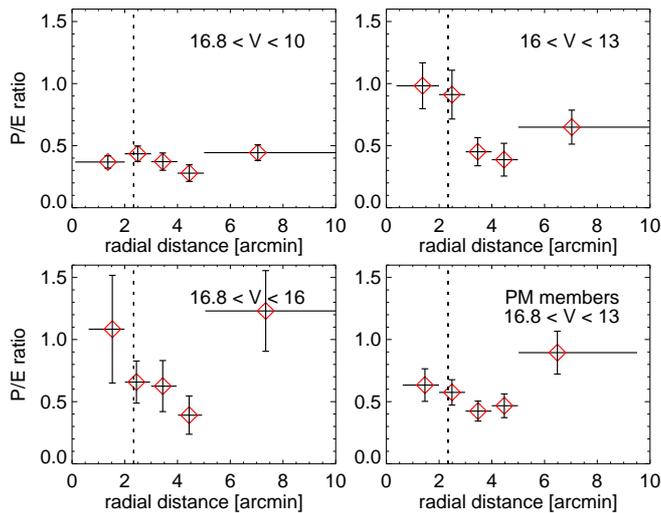}
\caption{Radial distributions of the ratio of primordial stars to extreme stars as function of the distance from the cluster center. The diamonds give the mean value of the radial bin whose range is indicated by the horizontal bar through the symbol. The dotted vertical line gives the half-mass radius of the cluster. }\label{Fig:rad dist}
\end{center}
\end{figure}

A Kolmogorov-Smirnov test indicates a 4.5\% probability that the radial distributions of the primordial (green curve) and the extreme populations (blue curve) are drawn from the same parent distribution. The largest difference between the primordial and extreme population is found at a distance of roughly 3 arcmin from the cluster center. We also divided the full sample into subgroups: RGB stars ($16<V<13$), SGB stars ($16.8<V<16$), and proper motion members ($V > 13$; proper motion data from \citealt{Zloczewski2012}). We divided each group into five bins of radial distance to the cluster center. For each bin we calculated the mean distance to the cluster center for each population, and computed their number ratios. The radial distribution of the ratio between primordial and extreme stars is given in Fig.~\ref{Fig:rad dist}. Overplotted are the uncertainties given by the Poisson statistic, $\sigma = N_P/N_E\sqrt{1/N_P + 1/N_E}$, where $N_P$ and $N_E$ represent, respectively, the number of primordial and extreme stars within each bin. The dotted line indicates the half-mass radius $r_h = 2.34$\,arcmin \citep{Harris1996}.

We compare our radial distribution with that of \citet{Kravtsov2011}. They claim a strong radial segregation between the primordial and extreme populations, based on ground-based broadband $UBVI$ photometry of RGB stars. They conclude that RGB stars redder in ($U-B$) are centrally concentrated within the $r_h$, while the radial distribution of RGB stars bluer in ($U-B$) peaks outside this region. This rather extreme bimodality is incompatible with our results, which merely indicate that the primordial population falls off faster than the extreme population between a radial distance of 2 and 5 arcmin (see e.g. left panel of Fig.~\ref{Fig:phot cumul}, and Fig.~\ref{Fig:rad dist}) (within our sample). We do not see a radial segregation around the half-mass radius of the cluster. Our results are instead compatible with those of \citet{Milone2013}, who analysed multi-band {\sl Hubble Space Telescope} (HST) photometry, indicating no significant radial trend in the relative numbers of the three stellar populations within 6 arcmin of the cluster center. \citet{Milone2010} executed a Kolmogorov-Smirnov test on the radial distributions of different photometric populations on the SGB, indicating a 4\% probability that the primordial and extreme population have the same radial distribution, which they considered far from conclusive.\\

Recently, \citet{Kravtsov2014} finalised the study of the NGC\,6752 broadband photometry from \citet{Kravtsov2011}, including the radial distribution of blue horizontal branch (BHB) stars. They found that the brighter BHB (BBHB) stars are more centrally concentrated than the faint (FBHB), thus suggesting nitrogen enhanced red giants as the progenitors of BBHB stars, on the basis of qualitatively similar radial distributions. The ratio of BBHB to FBHB stars of \citet[Fig.~11]{Kravtsov2014} exhibits a similar morphology to our primordial-to-extreme star ratios in Fig.~\ref{Fig:rad dist}, with a minimum appearing at $R \approx 5$\,arcmin. They find also that the nitrogen enhanced extreme stars, RGBs redder in $U-B$ and the $U$-faint SGBs, are centrally concentrated (essentially within $r_h$) while their blue and $U$-faint counterparts exhibit flatter radial distributions in the field. This is contrary to our results indicating smooth distributions of extreme stars, while the primordial stars appear more centrally concentrated, falling off more rapidly outside $r_h$. \\

Both observed radial distributions can be argued for in terms of build-up scenarios for the cluster. On the one hand, 1D hydrodynamical simulations show that second-generation stars are formed more centrally than first-generation stars. This follows from the assumption that second-generation stars are made up of the ejecta from first-generation AGB stars. These ejecta accumulate in a cooling flow into the cluster core where the second-generation stars then form. Follow-up N-body simulations show that a large fraction of first-generation stars is lost early on. This because the cluster expands and early mass loss due to first-generation supernovae strips of the outer layers populated by first-generation stars  \citep{Dercole2008,Vesperini2013}. During the dynamical evolutionary phase, both stellar generations mix and the ratio of second-to-first generation stars will settle into a morphology which is similar to the one observed by \citet{Kravtsov2014}. \\

On the other hand, we know that He-enhanced stars (i.e. second-generation stars) are slightly less massive than their He-normal counterparts (i.e. first-generation stars) when they evolve off the main sequence. The cluster will be driven toward equipartition of kinetic energy if one assumes that the dynamical evolutionary phase is dominated by two-body relaxation. As a result, the He-enhanced red giants can end up having a more extended distribution than He-normal ones after a Hubble time or several relaxation times. This is actually what we seem to find in our data sample: the primordial, i.e. He-normal, population falls off faster than their He-enhanced counterparts (see Fig.~\ref{Fig:rad dist}). There is no obvious explanation to reconcile both observations. We suggest further investigation of a consistent HST data set which extends to larger radial distance from the cluster center.

%#######################################################################################

%******************************************************************
%                               DISCUSSION
%******************************************************************

\section{Discussion}
We now return to the observed abundance trends and discuss their implications.
\begin{figure*}
\begin{center}
\includegraphics[]{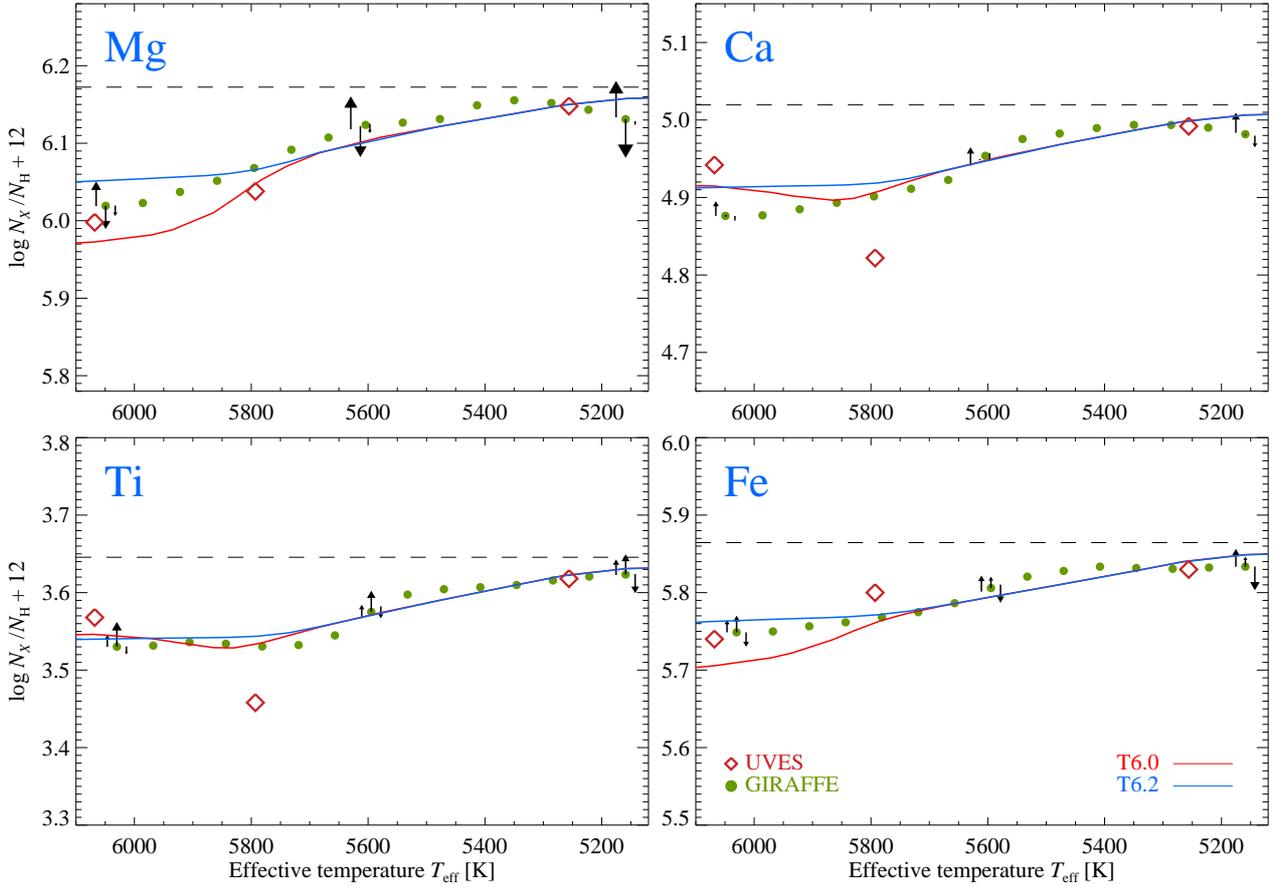}
\caption{Averaged abundance trends from Figure~\ref{Fig:abundances} (green bullets), compared to the predictions from stellar structure models including atomic diffusion with additional mixing with two different efficiencies, at an age of 13.5\,Gyr. Horizontal, dashed lines represent the initial abundances of the models, which have been adjusted so that predictions match the observed abundance level of the coolest stars. The groups of three arrows in each panel indicate the influence on averaged abundances from increasing the effective temperature by 40\,K (left), surface gravity by 0.1\,dex (middle) and microturbulence by 0.1\,\kms\ (right) -- see also Table~\ref{Tab:abund sensitivity}.}\label{Fig:AD-trends}
\end{center}
\end{figure*}

\subsection{Diffusion trends}\label{subsect:AD}
Figure~\ref{Fig:AD-trends} compares the averaged trends of magnesium, calcium, titanium and iron from Fig.~\ref{Fig:abundances} (bullets) to predictions from stellar evolution models computed at the cluster's metallicity $\FeH =-1.6$ (Paper~I). These models, introduced in Sect.\,\ref{sect:intro}, take into account the effects of atomic diffusion (AD), meaning radial transport of chemical elements, modelled from first principles, with the addition of a parametrised additional mixing (AddMix) of unknown physical origin. 

AddMix is parametrised as a function of temperature and density (i.e., depth), and its efficiency is set by a free parameter representing a rescaling to the efficiency at a reference temperature $T_0$. Figure~\ref{Fig:AD-trends} compares observations to AD model predictions at two different efficiencies of AddMix, where the model T6.0 ($\log T_0 = 6.0$) represents a low efficiency of AddMix, while T6.2 ($\log T_0 = 6.2$) represents high efficiency.
Dashed horizontal lines indicate the initial model composition, which has been slightly adjusted so that predictions match the abundances of the cooler stars, where models coincide. This in turn is an effect of the convective envelope reaching deeper into the star, until it surpasses the region affected by AddMix. 

Overall, we find good agreement between observations and predictions. The best correspondence is again found for the T6.2 model, as was the conclusion from the UVES analysis in Paper~I. 

In Sect.~\ref{sec:clusteranalysis} we argued that the chemical populations in NGC\,6752 differ in helium fraction by $\Delta Y \sim0.03$. \citet{Lind2011a} showed, however, that such a change in helium fraction has negligible influence on \Teff\ and \logg\ at a given evolutionary phase. We thus assume that the variation in $Y$ between stars of different populations does not influence the predictions from our stellar-structure models (cf. the contrasting case of NGC\,6397, \citealt{Korn2007}). We also note that we find no significant abundance difference between stars of different populations in Ca, Ti, Cr, Fe and Ni, except for the evolutionary trends discussed here. 
In the case of Mg, we find an insignificant tendency for the first-generation stars to exhibit a stronger trend than the second-generation stars, which is likely due to sampling of low-number statistics.
\\

The confirmation of the diffusion trends from Paper I brings further support to the assumption that AD is operational along the evolutionary sequence of NGC\,6752. Together with NGC\,6397 we now have two globular clusters in which effects of AD have been observed. Adding to this list the old, solar-metallicity open cluster M67, we can start to form a better picture about the r\^ole of AddMix and its behaviour with metallicity. Comparing the results in these three clusters with metallicities ranging from $\FeH = -2.1$ (NGC\,6397) over $-1.6$ (NGC\,6752) to $+0.06$ (M67, \citealt{Onehag2014}), we find a decreasing effect of AD. One might have to make a differentiation between the metal-poor clusters and the solar-like ones as the degree of Addmix is found to increase with metallicity, i.e. decreasing the efficiency of AD, for the metal-poor clusters while AddMix seems unnecessary to explain the abundance trends in M67. As a final note we would like to comment on yet another metal-poor cluster which was studied by \citet{Mucciarelli2011}, namely M4 (NGC\,6121) at $\FeH = -1.16$. Abundances of Fe and Li were derived for stars between the TOP and RGB. Although no trend in Fe was observed, the authors argue that the primordial lithium abundance (see Sect.~\ref{sec:evol-lithium}) can be reproduced assuming very efficient AddMix (T6.30). If this is true, the efficiency of AddMix is correlated with metallicity in the metal-poor regime. On the metal-rich side of the metallicity scale, however, AddMix does not seem to be required. One possible explanation is the fact that the outer convection zone extends further inside the star, surpassing the region where AddMix can operate. Continuing on this line of thought, one would expect the efficiency of AddMix to decrease with decreasing metallicity, as AddMix scales with the extent of the outer convection zone. Further analyses on globular clusters bridging the metallicity range between the metal-poor and the metal-rich regimes will shed light on the physical origin of AddMix. This is of utmost importance since any physical effects that alter the surface elemental composition of stars during their evolution must be taken into consideration in detailed studies of Galactic chemical evolution.

\subsection{Evolution of lithium} \label{sec:evol-lithium}

\begin{figure}
\begin{center}
\includegraphics[]{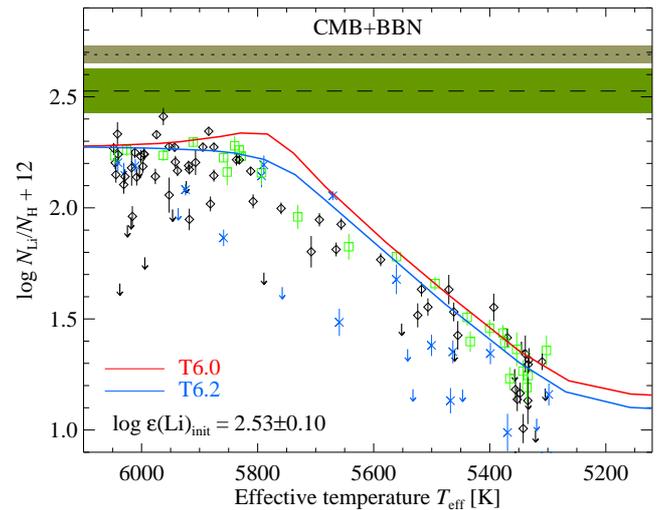}
\caption{Observed lithium abundances (coloured symbols as in Fig.~\ref{Fig:pop_abun}), compared to stellar evolution model predictions for two different efficiencies of AddMix (see text). 
The initial abundance of the models (horizontal dashed line and shaded region), $\abund{Li} = 2.53 \pm 0.10$, compares well to the predicted primordial lithium abundance (dotted horizontal line and shaded region), $\abund{Li} = 2.69 \pm 0.04$.}\label{Fig:lithium}
\end{center}
\end{figure}

Figure~\ref{Fig:lithium} compares our derived lithium abundances to model predictions, with stars identified by their chemical population membership as determined in Sect.~\ref{sec:clusteranalysis}. Compared to the traditional Spite plateau identified among field stars, surface lithium abundances of these GC stars are known to be dramatically affected both by internal and external processes. 
Firstly, surface layers of more evolved stars are diluted by essentially lithium-free processed material dredged up as the surface convection zone deepens along the subgiant branch. Secondly, star-to-star scatter is caused by processed material from short-lived massive stars mixing with the gas from which second-generation stars are forming, in what is known as intra-cluster pollution (see Sect.~\ref{sec:anticorr}). 
Even selecting the least evolved, least polluted stars of our sample, the same type of mostly non-destructive surface depletion by gravitational settling as has been discussed for other elements is at work for lithium (as in all Spite plateau stars). As a matter of fact, lithium is the metal affected the most by gravitational settling.\\

Selecting the least processed stars in the cluster is however not straightforward. Surface dilution sets in after the TOP at $T_\text{eff} \approx 5900$\,K, which happens to coincide with the limit below which our photometric identification technique becomes less reliable (see Sect.~\ref{sec:clusteranalysis}). Only four TOP stars (out of 43) are identified as belonging to the primordial population. Correcting their measured lithium abundances for the effect of atomic diffusion predicted by the T6.2 model ($\sim +0.25$\,dex), we find an average initial lithium abundance $\abund{Li} = 2.52 \pm 0.03$ (statistical error). 

\begin{figure*}[ht]
\begin{center}
\includegraphics[]{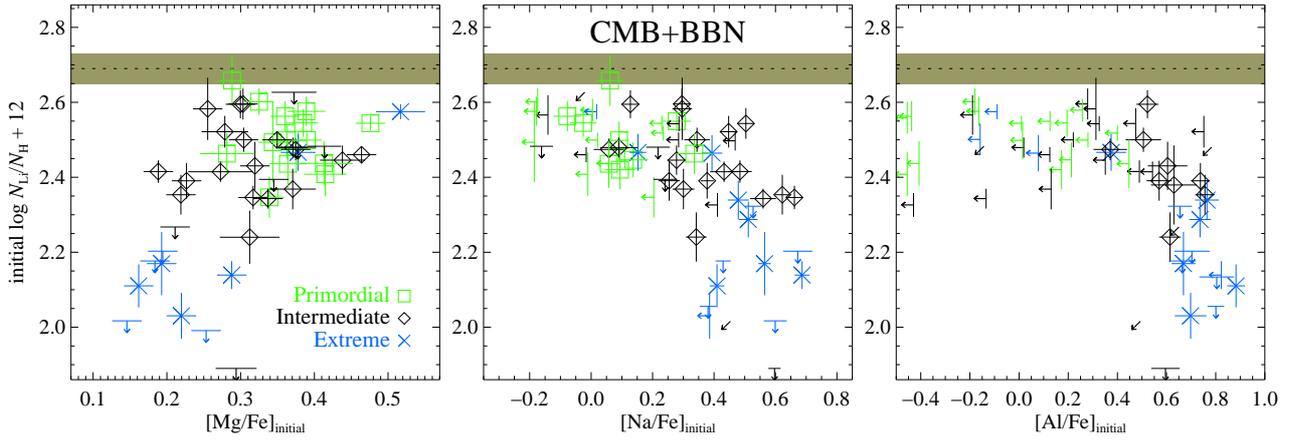}
\caption{Initial abundances (corrected for atomic diffusion and dredge-up) of lithium, compared to the anticorrelation elements magnesium (left), sodium (middle) and aluminium (right). The dotted line and shaded region refer to the predicted primordial abundance based on on BBN calibrated on the CMB (see text). The faintest ($V > 16.8$) stars have been excluded from the comparison. Different coloured symbols correspond to the chemical populations deduced from the cluster analysis.}\label{Fig:lithium_anticor}
\end{center}
\end{figure*}

Figure~\ref{Fig:pop_abun} indicates first-generation stars to be poor in sodium, with $\abund{Na} < 4.7$, in agreement with \citep{Carretta2013}. Selecting instead on this chemical criterion, matches 23 out of 30 TOP stars with observations of both sodium and lithium, indicating an initial abundance $\abund{Li} = 2.50 \pm 0.07$. 

Applying these two selection criteria separately to the full sample of stars, including those affected by dredge-up, indicates $\abund{Li} = 2.54 \pm 0.05$ (25 stars) and $\abund{Li} = 2.52 \pm 0.06$ (52 stars). The strongest constraint is found when combining the two criteria, giving $\abund{Li} = 2.53 \pm 0.05$ (21 stars). We adopt this as our best estimate of the initial lithium abundance, but enlarge the error bar to 0.10\,dex in accordance with the uncertainty in selection and the systematic uncertainties discussed in Sect.~\ref{sec:lithium}.
These results are in good agreement with Paper~I, where an initial level of $\abund{Li} = 2.58\pm0.10$ was deduced from two TOP stars.\\

Continuing this line of reasoning, we compare in Figure~\ref{Fig:lithium_anticor} the initial abundances of lithium to magnesium, sodium and aluminium, after correcting for evolutionary effects (AD and dredge-up).
While magnesium abundances appear smoothly correlated with lithium, both sodium and aluminium exhibit a two-zone behaviour. In both Figure~\ref{Fig:lithium} and \ref{Fig:lithium_anticor}, the predicted primordial Li abundance based on Big Bang Nucleosynthesis (BBN) calibrated on the cosmological background radiation (CMB) as observed by Planck, is given by the shaded area at  $\log\varepsilon$(Li)$_{\mbox{\scriptsize CMB+BBN}} = 2.69\pm0.04$ \citep{Coc2013}. The independent analysis of \citet{Steigman2013} prefers an insignificantly higher value of $2.72\pm0.04$.

\section{Conclusions}
Our differential 1D-LTE/NLTE based spectroscopic analysis of 194 stars observed at medium-high resolution, indicates weak ($\sim 0.1$\,dex) systematic abundance trends of heavy elements with evolutionary phase along the subgiant branch, in magnesium, calcium, titanium and iron. Although these trends are of low statistical significance taken individually, they are found to be in good agreement with those determined by independent methods in Paper~I, as well as predictions from stellar structure models including atomic diffusion with efficient additional mixing (the T6.2 model). 
The trends are not likely to be caused by systematic errors on, e.g., effective temperature, as flattening the abundance trend in iron would require an increase of 200\,K on the turnoff point, implying an unrealistically large error in $\Delta\text{\Teff}(\text{TOP}-\text{bRGB})$ of 25\,\%.\\

Abundances of lithium, sodium, magnesium, aluminium and silicon exhibit intrinsic (anti)correlated variations, indicating the presence of multiple stellar populations in the cluster. It has previously been shown that these can be identified -- predicted -- using $uvby$ Str\"omgren photometry of stars on the RGB \citep{Carretta2011}. We have here extended this work to fainter, less evolved stars on the subgiant branch down to $V \approx 16.8$, ($T_{\rm eff} \approx 5900$\,K) close to the TOP. It seems there is a genuine colour index for every population with a given abundance pattern in the sense that the polluted chemical composition of second-generation stars makes them appear redder than first-generation stars. We have investigated the radial distribution of the different stellar populations by use of $uvby$ Str\"omgren photometry. We find a weakly significant indication of a weak radial segregation where the center of the globular cluster is slightly overpopulated by the primordial population, compared to the outskirts. \\

Combining photometric and spectroscopic information, we have attempted to identify the least polluted (primordial) stars in the cluster. Correcting their abundance patterns for the expected effects of atomic diffusion, we predict an initial lithium abundance of $\abund{Li} = 2.53 \pm 0.10$. This is in good agreement with the results of Paper~I, $\abund{Li} = 2.58 \pm 0.10$, and similar to the initial composition of the less massive globular cluster NGC\,6397, $\abund{Li} = 2.57 \pm 0.10$ \citep{Nordlander2012}. For both clusters, the initial lithium abundance compares reasonably well with Planck-calibrated predictions of standard BBN, $\log\varepsilon$(Li)$_{\mbox{\scriptsize CMB+BBN}} = 2.69\pm0.04$ \citep{Coc2013}. \\

Lithium is but one element affected by atomic diffusion and additional mixing, as all elements are affected to varying degree by these physical processes.
With recent constraints on atomic diffusion in M67 \citep{Onehag2014}, it is now important to bridge the gap between metal-poor globular clusters and the regime of solar-metallicity stars.
This will help shed light on the physical nature of additional mixing such that surface abundances of late-type stars can systematically be corrected for atomic diffusion and their initial abundances reliably recovered, for the benefit of such diverse science cases as studies of planet-hosting stars or the chemical evolution of the Galaxy.

\begin{acknowledgements}
We thank O. Richard for providing stellar-structure models, F.~Grundahl for providing the $uvby$ Str\"omgren photometric data, and Y.~Osorio for computing NLTE corrections for magnesium.
PG and AK thank the European Science Foundation for support in the framework of EuroGENESIS. TN and AK acknowledge support by the Swedish National Space Board.
\end{acknowledgements}

\bibliographystyle{aa}
\bibliography{allreferences_giraffe}

\end{document}